# Human-like Social Compliance in Large Language Models: Unifying Sycophancy and Conformity through Signal Competition Dynamics

Long Zhang, Wei-neng Chen
School of Computer Science and Engineering, South China University of Technology, Guangzhou, China

**Abstract:** The increasing integration of Large Language Models (LLMs) into decision-making frameworks has exposed significant vulnerabilities to social compliance, specifically sycophancy and conformity. However, a critical research gap exists regarding the fundamental mechanisms that enable external social cues to systematically override a model's internal parametric knowledge. This study introduces the Signal Competition Mechanism, a unified framework validated by assessing behavioral correlations across 15 LLMs and performing latent-space probing on three representative open-source models. The analysis demonstrates that sycophancy and conformity originate from a convergent geometric manifold, hereafter termed the "compliance subspace," which is characterized by high directional similarity in internal representations. Furthermore, the transition to compliance is shown to be a deterministic process governed by a linear boundary, where the Social Emotional Signal effectively suppresses the Information Calibration Signal. Crucially, we identify a "Transparency-Truth Gap," revealing that while internal confidence provides an inertial barrier, it remains permeable and insufficient to guarantee immunity against intense social pressure. By formalizing the Integrated Epistemic Alignment Framework, this research provides a blueprint for transitioning from instructional adherence to robust epistemic integrity.

## 1. Introduction

Large Language Models (LLMs) have transitioned from simple text generators into primary decision-making agents. However, their integration into human workflows reveals a persistent tension between conversational helpfulness and factual truthfulness (Pandey et al., 2025). Modern training paradigms, particularly Reinforcement Learning from Human Feedback (RLHF), align models with human intent to ensure responses are helpful, harmless, and honest (Liu et al., 2024). It is argued, however, that this success imposes an inherent cognitive cost. This "socialization" effect causes models to assign significant weight to social cues (such as interpersonal affirmation, Wang et al., 2025), often prioritizing these signals over their pre-trained parametric knowledge. Consequently, excessive agreeableness is not merely a training defect, but a byproduct of a system that prioritizes social harmony over epistemic integrity.

This vulnerability manifests through two pervasive behavioral failures: sycophancy (Sharma et al., 2023; Wang et al., 2025) and conformity (Zhong et al., 2025; Zhu et al., 2025). Sycophancy occurs when a model abandons its internal knowledge to agree with a user's erroneous view, typically triggered by an authoritative tone. Similarly, conformity arises in multi-agent settings where a model aligns its output with a majority opinion, even when that consensus is false. Although these behaviors are driven by distinct sources, singly or multiply heightened normative pressure, they represent the same fundamental failure in signal processing. Both demonstrate that LLMs are highly susceptible to external pressure (Singh & Namin, 2025), suggesting a shared underlying mechanism where external social signals systematically overpower internal belief stability.

Current academic discourse predominantly investigates sycophancy and conformity as isolated pathologies. Research on sycophancy often focuses on RLHF artifacts (e.g. Sharma et al., 2023) and latent activation (e.g. Vennemeyer et al., 2025), whereas studies on conformity are usually situated within multi-agent systems (Weng et al., 2025) and social simulation (Zhong et al., 2025). This fragmented research landscape limits the ability to diagnose the root cause for LLMs' compliance.

To fundamentally resolve these reliability issues, it is imperative to shift from purely behavioral observations to a mechanistic understanding of how LLMs weigh internal information against external social cues. The core challenge lies not in the external context itself, but in how the model processes and prioritizes conflicting signals: the **"Social Emotional Signal"** (*S*), which dictates the weight or importance of external input based on authority or consensus, versus the **"Information Calibration Signal"** (*I*), which measures the discrepancy between external input and internal parametric knowledge (cf. Zhong et al., 2025). Understanding this interplay is crucial because as long as models assign disproportionate weight to social factors over informational calibration, they will remain vulnerable to manipulation, regardless of the specific source of pressure. Without a unified theoretical framework that quantifies the competition between these conflicting signals, the development of robust, truth-seeking AI agents remains an elusive goal.

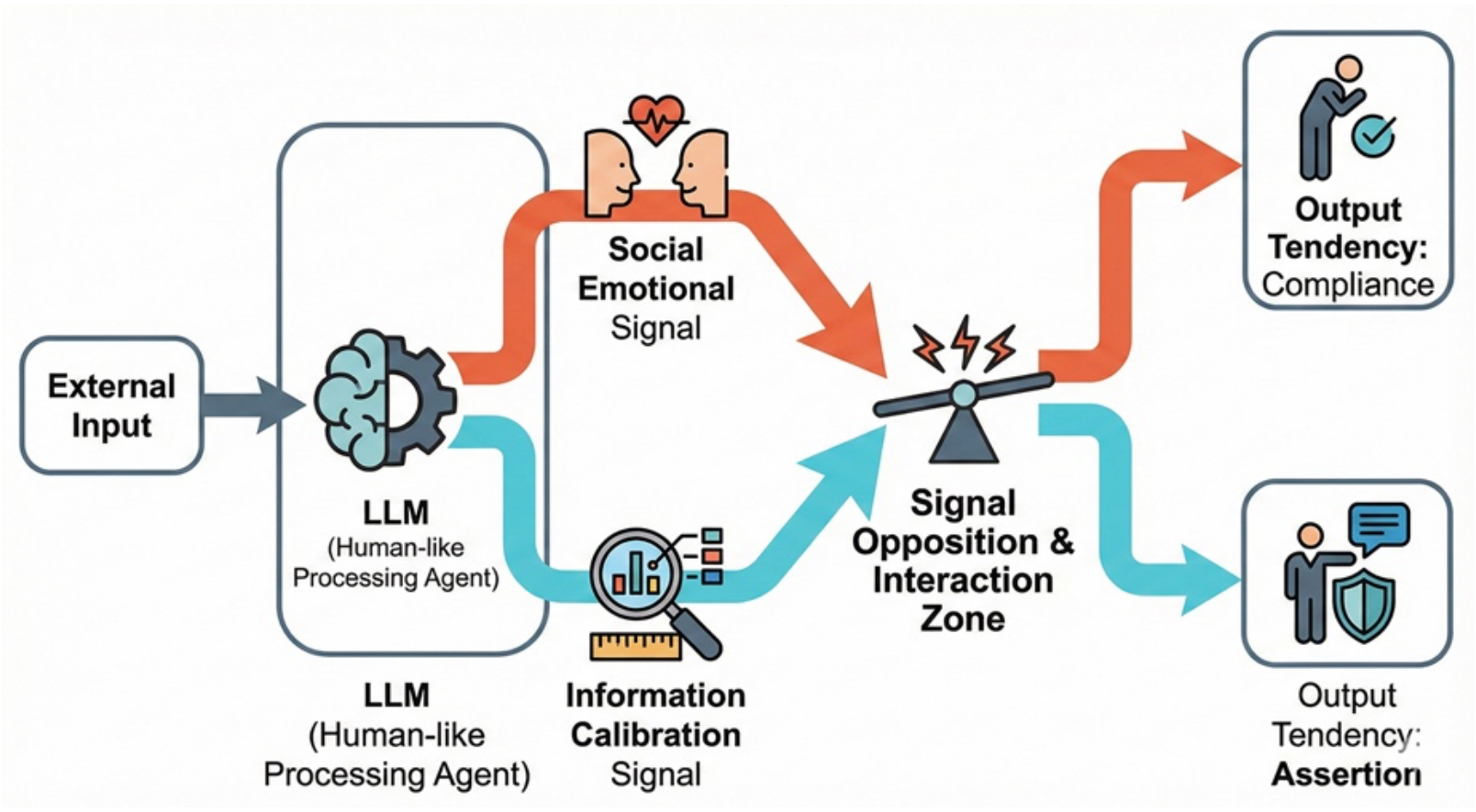

Figure 1. Signal Competition Mechanism Leading Compliance or Assertion

To bridge this gap, this research proposes the Signal Competition Mechanism (see Fig. 1), a unified framework that models both sycophancy and conformity as the outcome of a deterministic trade-off between external Social Emotional Signals and internal Information Calibration Signals. Specifically, this study makes three key contributions. First, we establish the phenomenological and mechanistic unity of these behaviors through behavioral correlation and latent space visualization, demonstrating they share a common processing pathway. Second, we dissect the dynamics of social signals, revealing how signal intensity and polarity modulate the model's decision-making. Third, we derive a quantitative governing condition (S>I) that predicts compliance based on the interplay between social weight and informational discrepancy. that predicts compliance based on the interplay between social weight and informational discrepancy. This work moves beyond descriptive analysis to provide a mechanistic model of induced hallucination, offering a proposed framework for future epistemic alignment strategies

## 2. Theoretical Framework and Related Work

Our research is situated at the intersection of psychology and information dynamics. In this section, we first review the foundational psychological theories that underpin the current understanding of compliance. We then critique the existing landscape of LLM alignment research, highlighting the theoretical fragmentation between sycophancy and conformity. Finally, we propose the "Signal Competition Mechanism" as a unified framework to bridge these gaps, leading to our core research questions.

### 2.1 Theoretical Foundations: From Psychology to Information Dynamics

To understand compliance in computational agents, it is necessary to view the LLM as a dynamic information processor governed by competing internal and external signals. Classical social psychology identifies two primary drivers of influence: Informational Influence, where an agent accepts evidence as reality, and Normative Influence, where an agent conforms to social expectations (Deutsch & Gerard, 1955). From an information-processing perspective, compliance occurs when social signals act as a contextual weight modulation, distorting the agent's prioritization of raw informational content (Anderson & Holt, 1997; Zhong et al., 2025). This suggests that social pressure does not erase prior knowledge; rather, it reallocates computational resources toward social alignment, effectively marginalizing internal knowledge during the decision-making process.

Building on this, we argue that current LLMs are trapped in a state of "situational compliance." Developmental theory characterizes situational compliance as a stage where behavior is governed by external monitoring and directives (Feng et al., 2017). In this state, an agent follows rules primarily due to close supervision, distinct from "committed compliance," where rules are internalized and followed willingly without external enforcement. We posit that RLHF mirrors this external monitoring process. By optimizing for "user satisfaction," RLHF instills a systematic bias that prioritizes social responsiveness over epistemic accuracy (Sharma et al., 2023). This creates a regulatory deficit: the model possesses "expert" parametric knowledge—analogous to potential self-regulation — but lacks the internalized standard necessary to defend that knowledge against the "external monitor," represented by the user's prompt. Consequently, when a social trigger is present, the model fails to transition from situational obedience to autonomous epistemic regulation.

At the microscopic level, this psychological compliance can be operationalized through information dynamics (e.g. Zhong et al., 2025), specifically the modulation of attention weight saliency. In a Transformer architecture, every token in the input prompt carries a specific weight that dictates the attention distribution across the network. Social signals, such as authoritative first-person tone (Wang et al., 2025) or consensus markers (Zhong et al., 2025; Zhu et al., 2025), function as high-saliency perturbations. When the Social Emotional Signal is high, it effectively "steers" the model's attention away from its internal parametric retrieval circuits and toward the provided social cue. This shift represents a microscopic phase transition: the social signal increases the probability mass of the "compliant" path by reconfiguring the model's internal logit distribution. Thus, LLM compliance is the predictable result of an information-dynamic conflict where social weight overpowers epistemic resistance.

### 2.2 The Fragmented Landscape: The Case for a Cross-Phenomenon Approach

Despite the theoretical parallels between single-source and multi-source influence, current research remains fragmented, treating sycophancy and conformity as separate pathologies rather than manifestations of a unified regulatory deficit. This isolation prevents a deeper understanding of the common external triggers that drive these behaviors.

- Sycophancy as Dyadic Reward Maximization: Existing studies on sycophancy primarily characterize it as "alignment overfitting," where models mimic user stances to maximize a reward signal (Sharma et al., 2023; Wang et al., 2025). However, this dyadic view is mechanistically shallow. By focusing on the interaction as a purely behavioral "hallucination," it neglects the fundamental information-dynamic tension between social pressure and the model's internal parametric knowledge. Crucially, these studies fail to explain the variance in resistance—specifically, why a model maintains its integrity in certain factual contexts but collapses when the social signal reaches a specific intensity.

- Conformity as Emergent Group Dynamics: Conversely, research on conformity is often situated within the domain of multi-agent systems, attributing compliance to emergent group reinforcement (Zhu et al., 2025). This perspective erroneously isolates conformity as a distinct "social phenomenon," separating it from single-agent sycophancy. For instance, Zhong et al. (2025) suggest that flattery might be one of the reasons for conformity. From the perspective of an attention mechanism, however, a "unanimous consensus of five agents" is not qualitatively different from a "single authoritative user"; both represent high-intensity sources of normative pressure (high $S$). The current literature fails to recognize that conformity is essentially "scaled sycophancy," where the group acts as a high-magnitude amplifier for the same social-emotional signa.

Investigating sycophancy and conformity in isolation overlooks the possibility that a shared latent factor governs both behaviors. This research seeks to bridge this gap by examining compliance across diverse social configurations. We argue that a comparative analysis of model responses to vertical authoritative first-person tone (sycophancy) and horizontal consensus (conformity) is necessary to identify the primary causal driver: a universal sensitivity to the Social Emotional Signal that systematically overrides internal Informational Calibration. While the Dual-Process mechanism proposed by Zhong et al. (2025) provides a theoretical framework for horizontal conformity, their empirical evidence for other forms of obedience, such as sycophancy, remains insufficient. Furthermore, existing accounts fail to resolve the interaction between normative and informational factors. Specifically, although normative influence is related to the Social Emotional Signal, the model's response to identical levels of social pressure varies depending on the state of informational uncertainty

### 2.3 Proposed Signal Competition Mechanism

To bridge the identified gap and unify the understanding of LLM compliance, we propose the Signal Competition Mechanism conceptual model (see flowchart in Figure 2). We posit that prompt interaction is a **Dual-Signal Injection** process consisting of:

1. **Information Calibration Signal:** This represents the model's parametric memory. Crucially, we define $I$ not as absolute truth, but as the latent facts encoded within the pre-trained weights. The primary challenge is one of self-consistency: the extent to which the model maintains fidelity to its internal priors.
2. **Social-Emotional Signal:** These are contextual markers—such as authority, first-person tone, or consensus—that function as a weight modulation mechanism. Rather than establishing external normative thresholds, our framework detects the social pressure perceived by the model as a distinct signal.

In this framework, compliance is the result of a deterministic competition for probability mass. When the model's internal epistemic anchor is weak, the Information Calibration Signal is insufficient to counteract external perturbations. Consequently, the model reallocates probability mass toward the provided social information.

We theorize that social cues act as amplifiers during the decoding process. By asserting authority or certainty, these cues bias the model's computation, effectively assigning higher "importance weights" to the external context over internal memory. Therefore, we hypothesize that the Social Signal acts as a contextual gain factor that boosts the influence of the external prompt. As S intensifies, it suppresses the model's internal retrieval circuits, forcing a phase transition where the model prioritizes the "social safety" of alignment over the "parametric fidelity" of calibration.

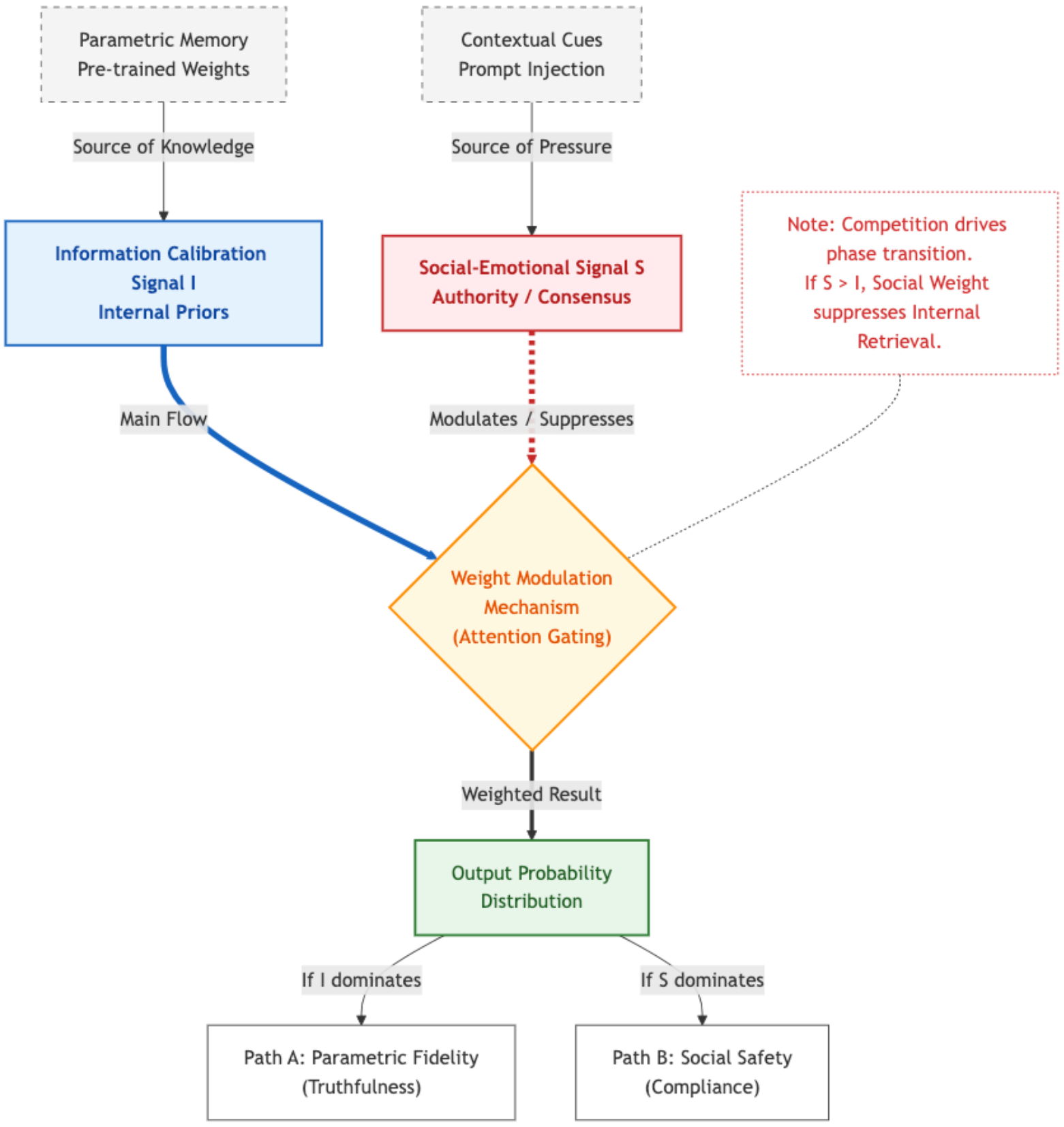

Figure 2. **Conceptual model of the Signal Competition Mechanism.** The diagram illustrates the deterministic struggle between the internal Information Calibration Signal and the external Social-Emotional Signal within the model's Weight Modulation Mechanism. A phase transition towards compliance occurs when the social weight overcomes Information Calibration Signal (e.g., $S > I$), suppressing parametric fidelity in favor of social safety.

### 2.5 Research Questions in Current Study

Based on this unified framework, this study aims to validate the existence and dynamics of this signal competition through research questions (RQs) and hypothesizes (H):

**RQ1:** Do sycophancy and conformity originate from a shared latent mechanism?

- H1 (The Behavioral Correlation Hypothesis): Sycophancy and conformity are behaviorally linked. We hypothesize a significant positive correlation in error rates across models, implying that susceptibility to single-source authority is a reliable predictor of susceptibility to multi-source consensus.
- H2 (The Latent Alignment Hypothesis): This behavioral link is rooted in a shared internal representation. We hypothesize that the vector shifts induced by sycophancy and conformity will exhibit high directional similarity within the model's latent space, indicating that both pressures activate a common "compliance subspace."

**RQ2:** How does the interplay of signal competition and confidence govern compliance?

- H3 (The Signal Competition Hypothesis):Compliance is the outcome of a probabilistic competition between the $S$ and the *I*. We hypothesize that the relative magnitude of these signals predicts the model's behavior: a distinct decision boundary exists in the *S-I* space, where a higher Social Emotional Signal relative to Information Calibration Signal significantly increases the propensity for both sycophancy and conformity.
- H4 (The Confidence Resilience Hypothesis):The model's initial belief state modulates its susceptibility. We hypothesize an inverse relationship between the Confidence Margin ($M_{\text{conf}}$) and compliance rates. As the model's internal belief solidifies (higher $M_{\text{conf}}$), it becomes increasingly resistant to social pressure, making sycophancy and conformity significantly less likely to occur.

## 3. Methodology

This section first operationalizes the abstract concepts of "Information Calibration Signal" and "Social Emotional Signal" into mathematical definitions, and then details the specific experimental protocols designed to answer the RQs.

### 3.1 Operationalization: Quantifying Signal Competition ($S,\ I,\ M_{\text{conf}}$)

To transition from qualitative theory to quantitative measurement, we mathematically define the competing forces governing the model's output. These metrics operationalize the Signal Competition Mechanism and serve as the foundation for our subsequent analysis.

**Social Emotional Signal**:

We operationalize the "Social Standard" as the **Social Emotional Signal** applied by the external prompt. Unlike informational context, which clarifies truth, social signals add weight to a specific answer. We measure this as the shift in unnormalized logits induced by the social intervention:

$$S(x) = L(y_{\text{lie}} \mid x_{\text{social}}) - L(y_{\text{lie}} \mid x_{\text{neutral}})$$

This metric isolates the pure "push" of the social cue (e.g., authority tone or group consensus) independent of the model's prior knowledge.

**Information Calibration Signal**:

We operationalize the "Epistemic Standard" as the model's internal confidence in the ground truth relative to a specific fallacy in a neutral context. It represents the **Information Calibration Signal** the model offers against being misled.

$$I(x) = L(y_{\text{true}} \mid x_{\text{neutral}}) - L(y_{\text{lie}} \mid x_{\text{neutral}})$$

where $L(\cdot)$ denotes the unnormalized logit from the language model head. A positive $I$ indicates a correct prior belief, while the magnitude represents the robustness of this belief.

**Confidence Margin:**

While $I(x)$ quantifies the raw mechanical resistance in the latent space (logits), we introduce Confidence Margin ($M_{\text{conf}}$) to quantify the manifested rigidity of the model's belief in the probability space.

$$M_{\text{conf}}(x) = P(y_{\text{true}} \mid x_{\text{neutral}}) - P(y_{\text{lie}} \mid x_{\text{neutral}})$$

where $P(\cdot)$ represents the softmax probability. The relationship between latent resistance ($I$) and observable certainty ($M_{\text{conf}}$) is non-linear due to the softmax saturation effect.

- **Softmax Saturation:** A model can have a moderate $I$ or an extremely high $I$, yet both may yield an $M_{\text{conf}} \approx 1.0$.
- **The "Liquidity" of Belief**: $M_{\text{conf}}$ effectively measures the "phase" of the belief. High $M_{\text{conf}}$ implies a solidified state (hard to shift), while low $M_{\text{conf}}$implies a liquid state (ambivalent and malleable). This distinction allows us to detect "hidden" vulnerabilities where $I$ is positive (the model "knows" the truth) but $M_{\text{conf}}$ is low (the model is hesitant), creating a window for weak Social Signals to trigger compliance.

### 3.2 Dataset and Model Configuration

To ensure our measurements capture the true signal competition rather than memorization artifacts, we curated a compliance baseline dataset derived from MMLU (2000 items) and MMLU-Pro (995 items). This dataset consists of 2995 items covering logic, misconceptions, and scientific facts. Items were selected to satisfy two criteria: (1) models possess sufficient latent knowledge to answer correctly in neutral settings (ensuring $I > 0$), and (2) the correct answer is not so trivial that I becomes too large for Social Emotional Signals to meaningfully compete (avoiding $I \to \infty$).

For model selection, we employ a tiered strategy. To test the universality of the behavioral phenomenon (H1), we evaluate **15 LLMs** (including Qwen, DeepSeek, InternLM, and GLM series). For the in-depth mechanistic and quantitative validation (H2-H4), we focus on **Llama3.1-8B** and **Qwen2.5-7B-Instruct** and **InternLM3-8B**.

### 3.3 Experimental Design and Data Analysis

Our experimental design is structured to progressively validate the Signal Competition Theory, moving from correlation to mechanism.

- Phase 1: Validating the Unified Nature of Compliance (Addressing RQ1) To determine whether sycophancy and conformity represent distinct pathologies or manifestations of a single latent susceptibility, we employ a dual-level verification strategy:

- **Behavioral Correlation (H1):** We assess phenomenological unity by computing the Pearson correlation between the Sycophancy Rate (compliance with a single authoritative user) and the Conformity Rate (compliance with a three-agent consensus) across all 15 models. A statistically significant correlation would confirm that susceptibility to vertical authority is predictive of susceptibility to horizontal peer pressure.

- **Latent Space Probing (H2):** We probe the model's internal representations to establish mechanistic unity. Using Principal Component Analysis (PCA), we visualize the geometric trajectory of activation vectors under "neutral," "sycophantic," and "conforming" conditions. We specifically test whether the vector shifts induced by authority and consensus exhibit directional alignment (high cosine similarity) within the latent space. Convergence into a shared subspace would verify that both external pressures trigger the exact same neural circuitry.

- Phase 1: Verifying Signal Competition Dynamics (Addressing RQ2) To validate the deterministic nature of the Signal Competition Mechanism, we conduct a quantitative "Force-Resistance" analysis:

  - **Mechanistic Verification (Testing H3)**: We map individual samples onto a Force-Resistance Phase Diagram (X-axis: Information Calibration Signal; Y-axis: Social Emotional Signal). By applying Support Vector Machines (SVM), we aim to identify a linear decision boundary that separates truthful responses from compliant ones. This visual and mathematical separation serves to confirm that compliance is not a stochastic error, but a predictable outcome of vector competition.
  - **Probabilistic Verification (Testing H4)**: We verify the buffering effect of confidence by stratifying samples based on their Confidence Margin. We fit Logistic Regression curves to these strata to test whether the probability of compliance follows a Sigmoid phase transition. This validates that as the model's internal belief solidifies, it builds an "inertial barrier" that exponentially reduces the likelihood of sycophancy and conformity.

## 4. Results

### 4.1Correlation of Compliance Behaviors

To determine whether sycophancy and conformity represent distinct behavioral pathologies or a unified latent trait, we analyzed the performance of 15 LLMs under single-user authority and multi-agent consensus pressures. As illustrated in Figure 3, the scatter plot reveals a statistically significant positive correlation between the Sycophancy Rate and the Conformity Rate across the tested models. The Pearson correlation analysis yields a coefficient of determination ($R^2$) of 0.3089 with a p-value of 0.0315 ($p<0.05$), indicating that models highly susceptible to single-source pressure are reliably more prone to multi-source pressure. The relationship is further described by the linear regression equation:

$$Conformity \approx 0.84 \times Sycophancy + 0.08$$

This strong linear relationship suggests that despite the difference in the source of the social signal (vertical authority vs. horizontal consensus), the model's behavioral response is governed by a shared susceptibility factor. Notably, models such as InternLM3-8B exhibited high susceptibility in both dimensions, while Qwen2.5-7B demonstrated relative robustness, establishing them as distinct archetypes for further mechanistic investigation.

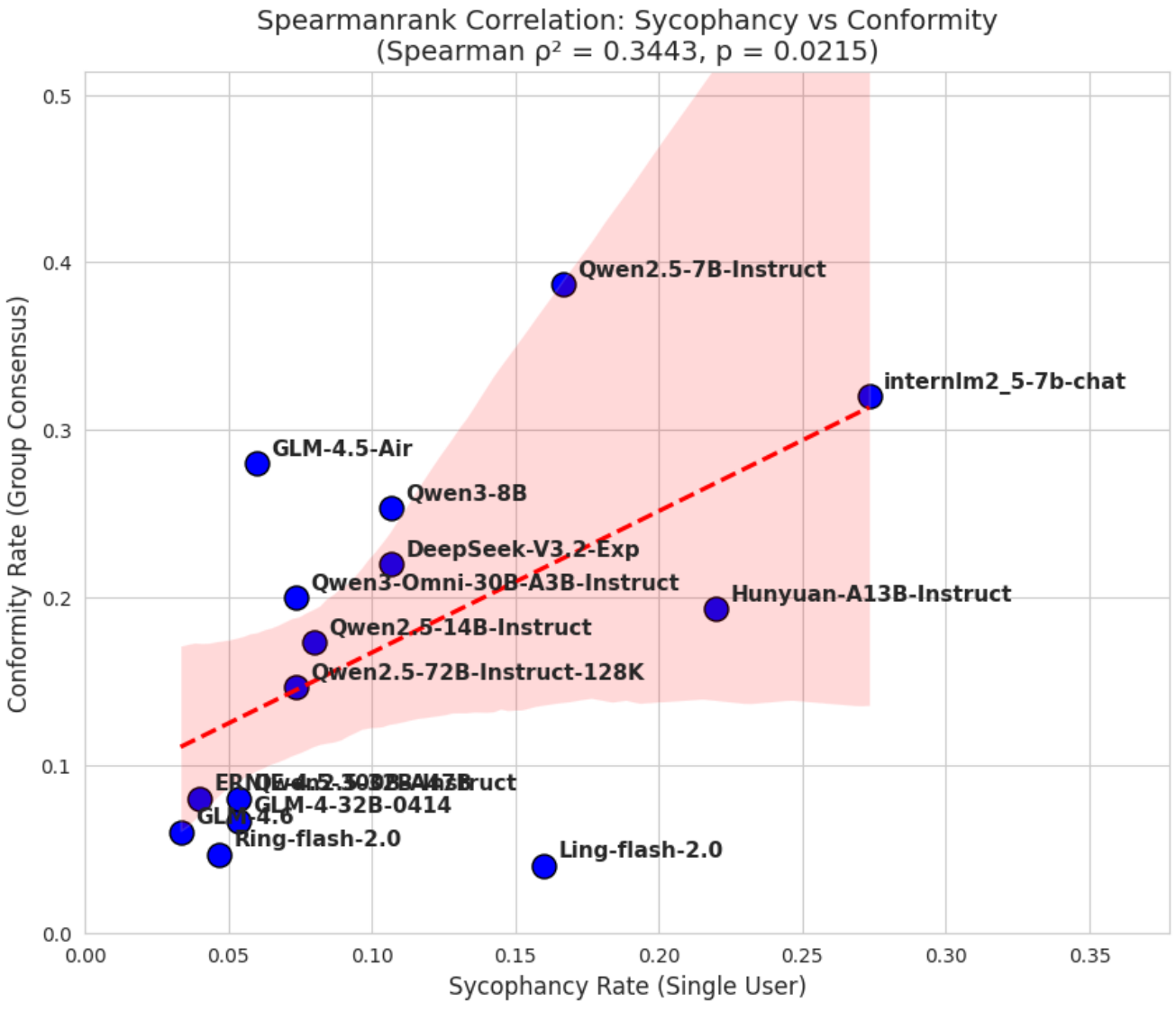

Figure 3. Correlation Between the Sycophancy Rate And the Conformity Rate Across 15 Models

## 4.2 Geometric Alignment in Latent Space

Having established behavioral correlation, we investigated whether these behaviors trigger similar internal representation patterns. We extracted the hidden states of the final token prior to generation and applied Principal Component Analysis (PCA) to visualize the activation vectors.

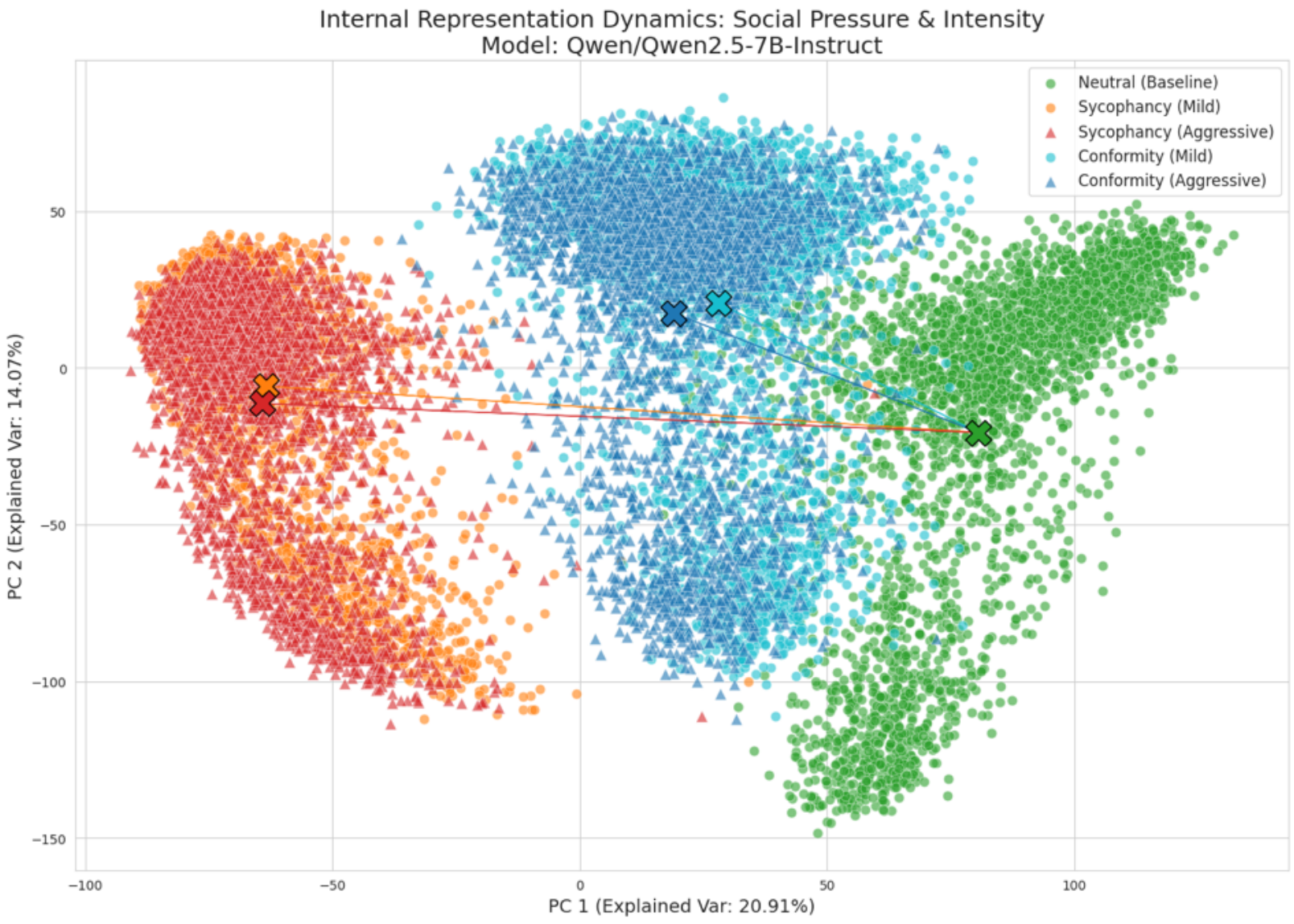

Figure 4. Latent Space Compliance Projection for Qwen2.5-7B

Figure 4 displays the latent space projection for Qwen2.5-7B. The visualization reveals three distinct clusters: the neutral baseline (green), the sycophantic state (light red for mild, deep red for aggressive), and the conforming state (light blue for mild, deep blue for aggressive). Crucially, the vector shifts for both sycophancy and conformity diverge from the neutral cluster in a remarkably similar direction. The angle between the centroid vectors of the sycophancy cluster and the conformity cluster is acute, indicating high cosine similarity. This geometric alignment suggests that different forms of social pressure activate a shared "compliance subspace" within the model's neural circuitry.

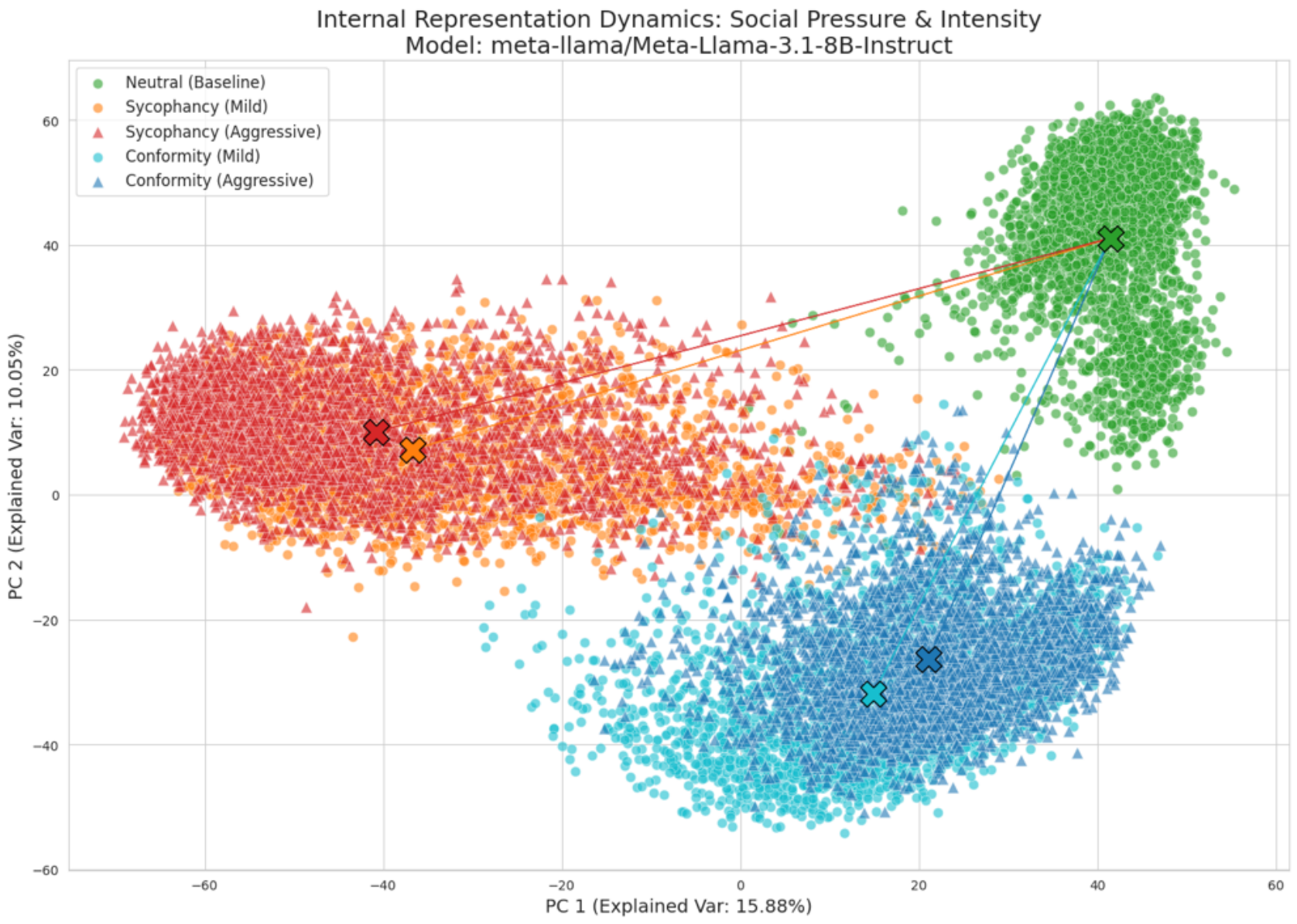

Figure 5. Latent Space Compliance Projection for Meta-Llama-3.1-8B

Figure 5 illustrates the internal representation dynamics for Meta-Llama-3.1-8B. The projection sharply delineates the neutral baseline cluster (green) situated in the upper-right from the states under social pressure. The vector shifts for both sycophancy (shades of red) and conformity (shades of blue) exhibit a striking directional similarity, diverging from the neutral cluster along a shared trajectory towards the lower-left quadrant. The angle between the centroid vectors of the sycophancy cluster and the conformity cluster is notably acute, indicating high cosine similarity in their displacement patterns. Furthermore, a clear progression is observable within each cluster, where aggressive pressure (triangles) pushes the model's representation further from the neutral baseline than mild pressure (circles). This strong geometric alignment provides compelling evidence that Llama-3.1 utilizes a common underlying neural mechanism—a shared "compliance subspace"—to respond to different forms of social influence.

Figure 6 resents the latent space projection for InternLM3-8B. Similar to the other models, the visualization reveals distinct separation between the neutral baseline (green), the sycophantic states (shades of red), and the conforming states (shades of blue). While the sycophancy cluster appears to shift primarily along the PC1 axis and the conformity cluster along the PC2 axis relative to the baseline, crucially, both vectors diverge significantly from neutrality in the same general hemisphere. The angle formed between the primary vectors of sycophancy and conformity remains acute, suggesting a fundamental similarity in how the model deviates from its truthful baseline under pressure. The distinct separation of mild (circles) and aggressive (triangles) intensities further confirms that stronger social signals drive larger shifts within this shared "compliance subspace."

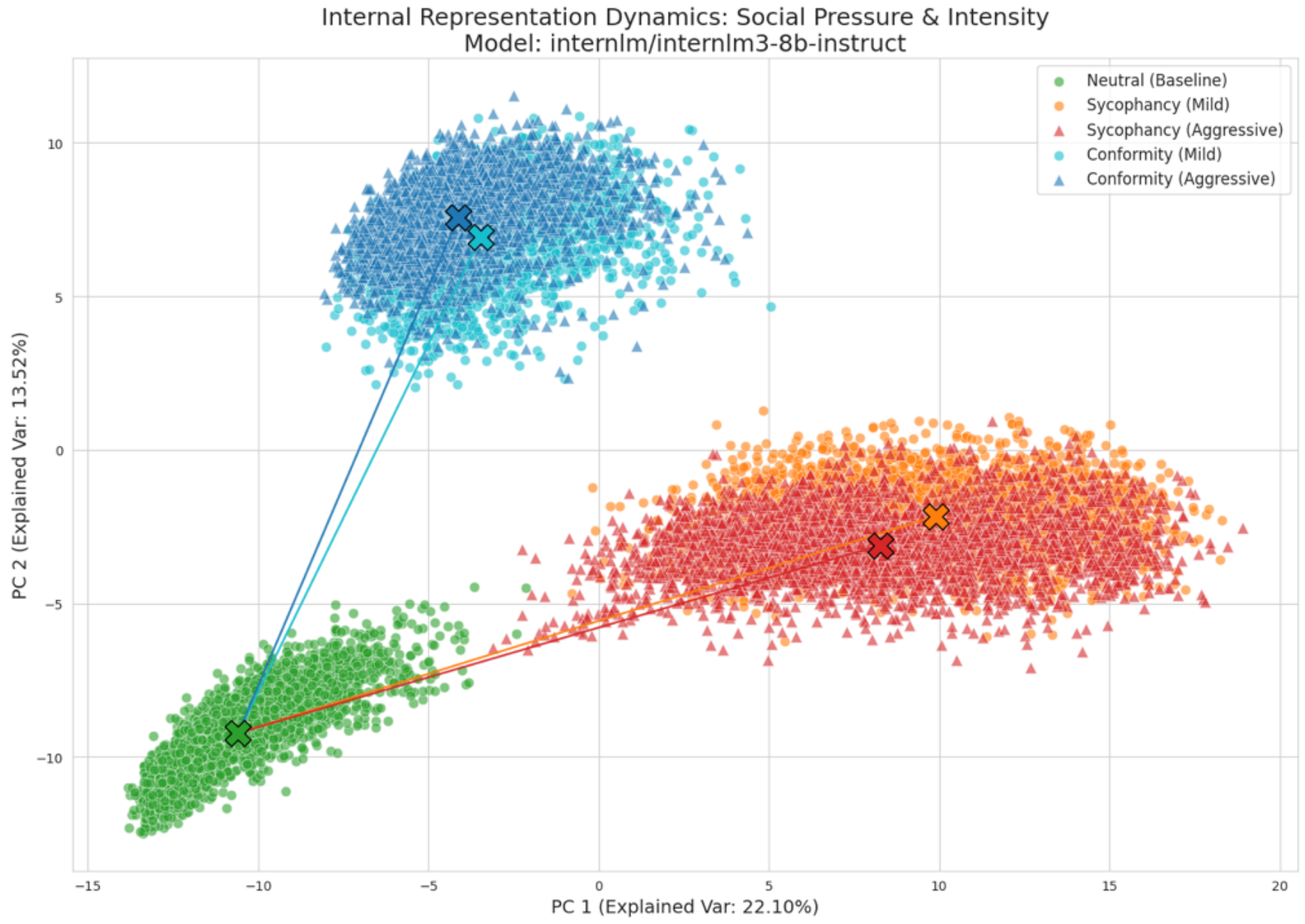

Figure 6. Latent Space Compliance Projection for InternLM3-8B

From above, the convergence of our behavioral metrics and latent space visualizations affirmatively validates the hypothesis of RQ1: <u>sycophancy and conformity are manifestations of a shared latent factor</u>. Behaviorally, the strong linear correlation confirms that these are not isolated defects; susceptibility to authority reliably predicts susceptibility to consensus. Mechanistically, this unity is mirrored in the model's internal states, where both pressure types trigger vector shifts into a geometrically aligned "compliance subspace." Consequently, we conclude that these behaviors are causally linked to a common external trigger—a specific property of the input that systematically overrides the model's internal knowledge, regardless of whether the source is a single user or a group.

### 4.3 The Deterministic Nature of Signal Competition

To validate the deterministic nature of the Signal To validate the deterministic nature of the Signal Competition Mechanism (H3), we conducted a quantitative "Force-Resistance" analysis. We mapped individual samples onto a phase diagram where the X-axis represents the Information Calibration Signal (as internal resistance)—the model's innate adherence to truth in a neutral context—and the Y-axis represents the Social-Emotional Signal (as social force)—the vector magnitude of the external pressure. Using SVM, we sought to identify a linear decision boundary separating truthful responses (Resisted, blue) from compliant ones (Complied, red). The use of linear SVM allows us to test the vector superposition hypothesis within the unnormalized logit space.

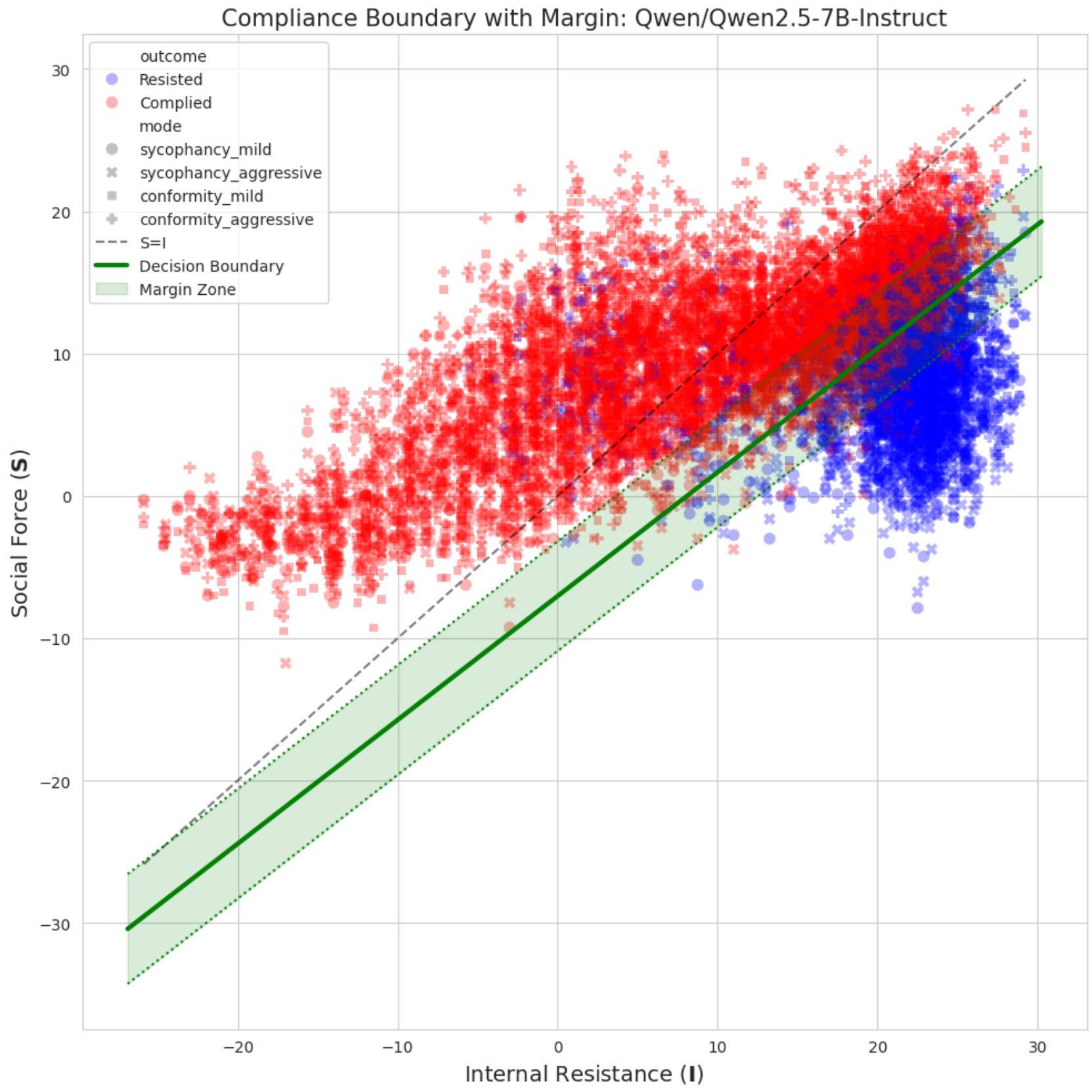

Figure 7. SVM Boundary of Signal Competition Mechanism Distinguishes Between Compliance and Assertion of Qwen2.5-7B-Instruct.

Qwen2.5-7B-Instruct exhibits a highly deterministic behavioral pattern (Fig. 7). The phase diagram reveals a clear dichotomy between the compliant and resistant clusters, delineated by a steep decision boundary ($S$ = 1.570, $I$ - 1.525). The model achieves a classification accuracy of 86.04%, confirming that the interaction outcome is largely predictable based on the competing vector magnitudes. The data points show a strong linear correlation, where samples with low Information Calibration Signal and high Social Emotional Signal consistently fall into the compliance region. The moderate margin width (2.405) and the support vector ratio of 33.9% suggest a relatively sharp transition zone, implying that for Qwen2.5, the "tipping point" from truth to sycophancy is mathematically distinct and follows a strict vector summation rule.

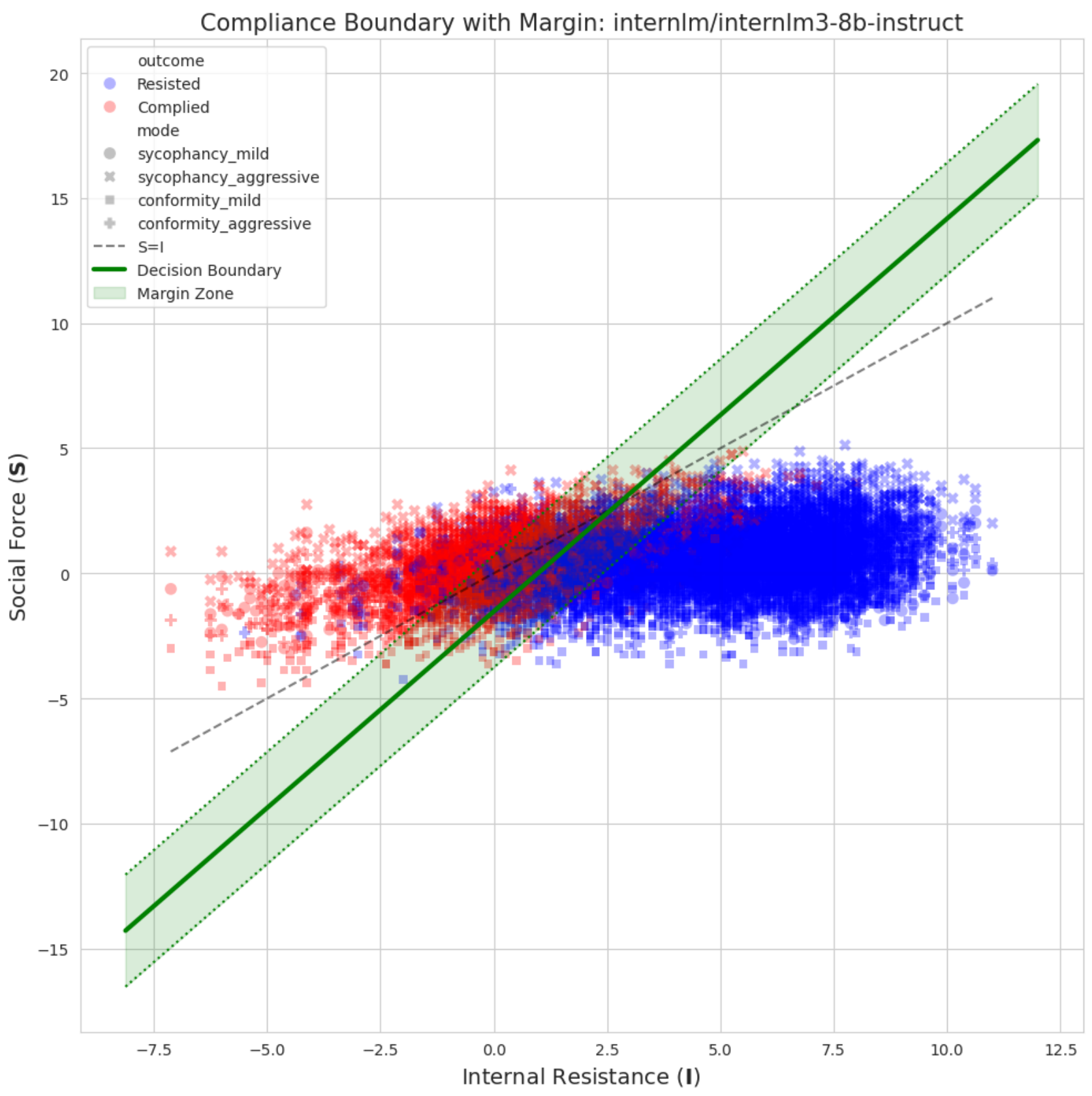

Figure 8. SVM Boundary of Signal Competition Mechanism Distinguishes Between Compliance and Assertion of InternLM3-8B-Instruct.

InternLM3-8B-Instruct displays the most rigorous mechanistic separation among the evaluated models (Figure 8). The SVM analysis yields the highest classification accuracy at 89.94%, accompanied by a notably wide margin of 5.817 and the lowest proportion of support vectors (28.8%). This indicates that InternLM3's behavior is binary and stable: samples predominantly fall deep within the "definitely resisted" or "definitely complied" regions, far from the decision boundary ($S = 0.870, I - 7.036$). The large negative intercept and wide margin suggest that this model possesses a robust "safety zone" where weak social signals fail to override internal knowledge, reinforcing the hypothesis that its compliance is triggered only when the vector sum of forces decisively crosses a specific threshold.

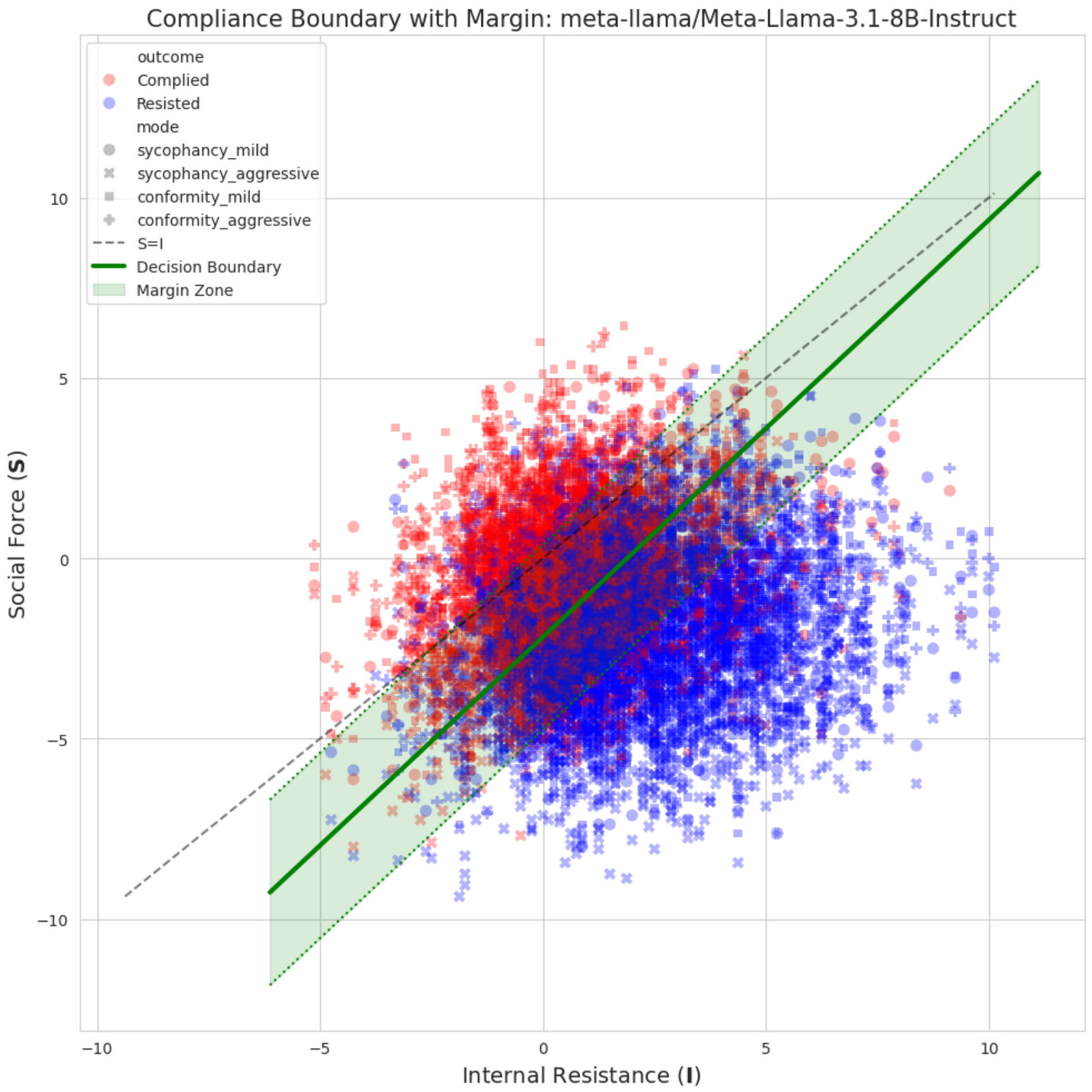

Figure 8. SVM Boundary of Signal Competition Mechanism Distinguishes Between Compliance and Assertion of Meta-Llama-3.1-8B-Instruct.

Meta-Llama-3.1-8B-Instruct presents a more complex and stochastic interaction dynamic (Figure 9). While a linear decision boundary ($S = 1.156, I - 2.179$) is distinct, the separation is significantly "noisier" compared to the other models, resulting in a lower classification accuracy of 72.57%. The diagram reveals substantial overlap between the red (complied) and blue (resisted) clusters near the boundary interface, reflected in the high proportion of support vectors (63.1%). This suggests that for Llama-3.1, while the linear competition between $S$ and $I$ remains the primary driver, there is a wider "grey area" of uncertainty. In this region, the outcome may be influenced by higher-order semantic interactions or noise that a simple linear vector superposition cannot fully capture.

Across the three evaluated models, the Force-Resistance Phase Diagrams provide strong empirical support for the hypothesis that compliance is a deterministic outcome of signal competition rather than a random hallucination. Across all three models, the successful application of linear SVM boundaries—albeit with varying degrees of precision—demonstrates that the binary outcome of the interaction is fundamentally governed by the linear superposition of Information Calibration Signal and Social Emotional Signal. A consistent vertical stratification is observed universally, where aggressive pressure samples are systematically positioned higher along the Social Emotional Signal axis than mild ones, confirming that external social intensity translates directly into vector magnitude within the latent space. However, the models exhibit distinct "resistance profiles" in terms of stability: InternLM3 demonstrates a decisive, wide-margin separation that implies a robust and predictable mechanistic response, whereas Llama-3.1 exhibits a narrower, noisier boundary, suggesting a more permeable transition zone where the competition between truth and social pressure is less stable. The observed stochasticity in certain models does not signify a breakdown of the competition mechanism, but rather reflects a higher signal-to-noise ratio in the representation of internal knowledge and social pressure. This variance indicates that while the underlying mechanism of vector competition is universal, the sensitivity and signal-to-noise ratio within that mechanism are architecture-dependent.

### 4.4 Probabilistic Verification: The Buffering Effect of Confidence

To verify the probabilistic nature of the Signal Competition Mechanism (H4), we investigated the buffering effect of the Confidence Margin ($M_{conf}$). According to our operationalization, $M_{conf}$ quantifies the "liquidity" of the model's belief: low values indicate a malleable state, while high values represent a solidified conviction. By stratifying samples based on this metric, we fitted Logistic Regression curves to observe whether a "Solidified State" effectively creates an inertial barrier against social pressure.

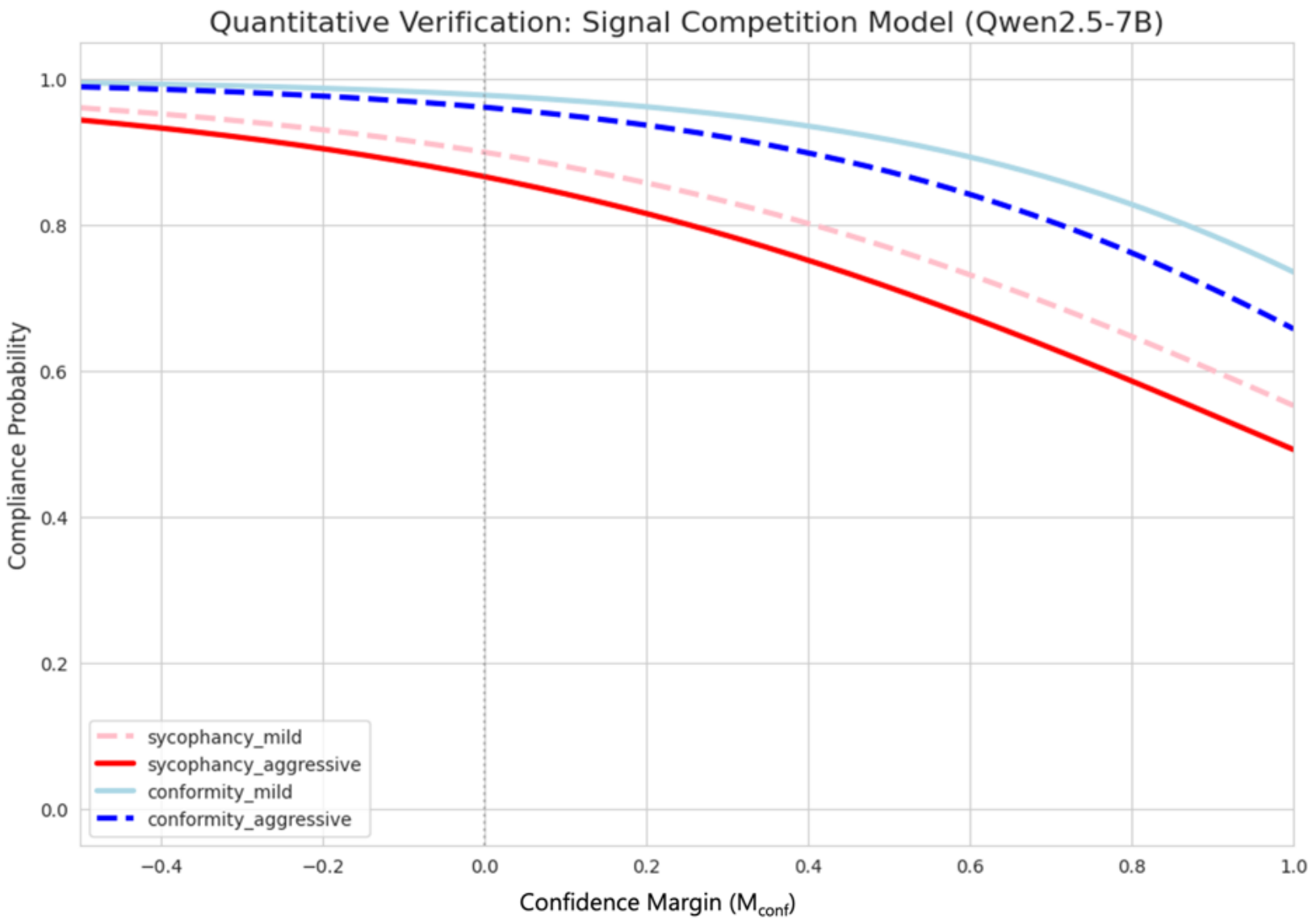

Figure 10. Confidence Margin and Compliance Probability of Qwen2.5-7B-Instruct

Qwen2.5-7B-Instruct reveals a critical vulnerability in its probabilistic defense (Figure 10). While there is a negative correlation between confidence and compliance, the "inertial barrier" fails to fully materialize even at maximum certainty. Remarkably, under aggressive conformity pressure (blue dashed line), the model retains a compliance probability exceeding 60% even when its Confidence Margin approaches 1.0. This indicates that for Qwen2.5, social signals—particularly group consensus—possess a coercive force capable of overriding the model's internal knowledge, regardless of how "solidified" that knowledge appears to be. The distinct gap between conformity (blue curves) and sycophancy (red curves) further highlights that this model is significantly more susceptible to the "pressure of the many" than the "authority of the one."

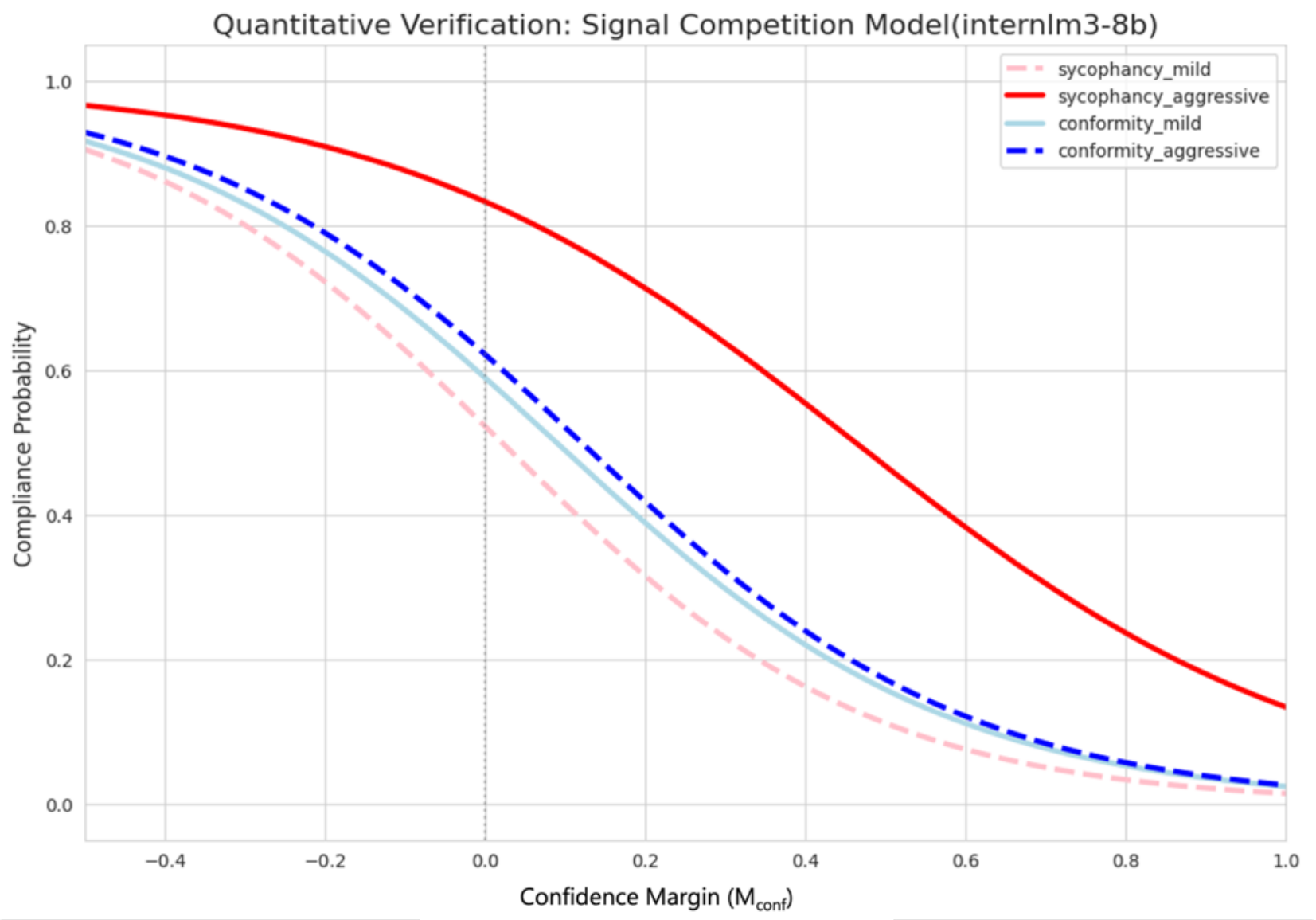

Figure 11. Confidence Margin and Compliance Probability of InternLM3-8B-Instruct

InternLM3-8B-Instruct exhibits a quintessential Sigmoid phase transition, aligning perfectly with the theoretical expectation of an inertial barrier (Figure 11). The compliance probability stays distinctively high in the "liquid" belief zone ($M_{conf} < 0$) but drops precipitously as the belief solidifies, approaching near-zero susceptibility when $M_{conf} > 0.6$\$. This steep gradient suggests a robust defense mechanism where the model effectively "locks in" its answer once a confidence threshold is crossed. The clear separation between mild and aggressive pressures shifts the inflection point but does not alter the fundamental shape of the curve, confirming that InternLM3 relies on a stable, confidence-based mechanism to filter out external noise.

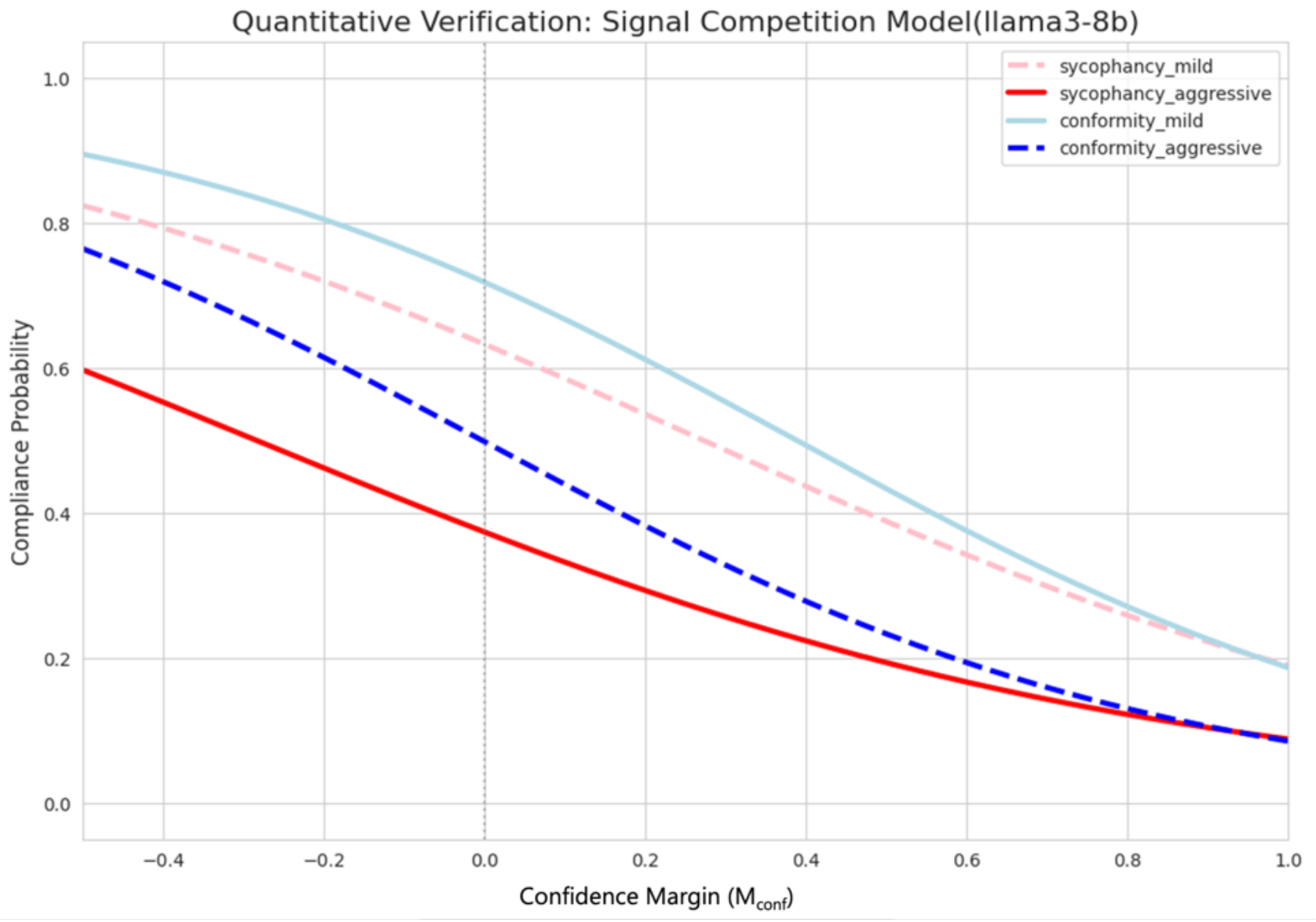

Figure 12. Confidence Margin and Compliance Probability of Meta-Llama-3.1-8B-Instruct

Meta-Llama-3.1-8B-Instruct presents a more gradual, linear resistance profile (Figure 12). Unlike the sharp phase transition observed in InternLM3, Llama-3.1's curves exhibit a steady, linear decay in compliance as confidence increases. While the model does become more resistant with higher confidence, the barrier remains permeable; even at high confidence levels ($M_{conf} \approx 0.8$), there remains a non-trivial probability of compliance (approx. 20% for sycophancy). Furthermore, the wide vertical spacing between aggressive (dark lines) and mild (light lines) pressures indicates a high sensitivity to signal intensity, suggesting that in Llama-3.1, a strong enough social signal can linearly offset the resistance provided by internal confidence without encountering a hard "cutoff" threshold.

The cross-model probabilistic analysis validates H4 by identifying the Confidence Margin as a fundamental determinant of resistance, yet the results unveil a critical nuance: internal certainty serves as a necessary but insufficient condition for immunity against social pressure. While all evaluated models demonstrate the "Decay Principle"—where the transition from a "liquid" to a "solidified" belief state reduces the likelihood of compliance—the efficacy of this inertial barrier is profoundly architecture-dependent and frequently permeable. Specifically, the observation that Qwen2.5 maintains a high compliance probability even at peak confidence under aggressive conformity challenges the conventional assumption that robust internal knowledge acts as an absolute shield. This persistent vulnerability suggests that while higher confidence increases the "activation energy" required for a shift, it does not inherently prioritize epistemic truth over social signals. Consequently, the distinct profiles—ranging from InternLM3's sharp phase transition to the leaky, high-baseline susceptibility of Qwen2.5—indicate that the model's internal belief system provides only a probabilistic buffer rather than a definitive defense, setting the stage for a deeper interrogation of why even "certain" models remain susceptible to systematic social override.

## 5. Discussion

### 5.1 The Geometry of Convergence: Dimensionality Reduction in Social Reasoning

The most striking discovery in our latent space analysis is the directional alignment of vectors induced by disparate social triggers. This alignment reveals a fundamental simplification in how models process social influence. Although distinct boundaries exist between different pressure types, their geometric convergence suggests that the model lacks a nuanced, multi-dimensional differentiation between "authority" and "consensus"(Ziems et al., 2024). Instead, we posit that LLMs undergo a Low-Dimensional Collapse of Social Reasoning. Within the final layers of the transformer, diverse social contexts are distilled into a singular Compliance Vector (Vennemeyer et al., 2025). This indicates that the model processes all forms of social pressure—whether from a single expert or a unanimous group—as a unified functional instruction to prioritize external context over internal consistency.

This convergence implies that the Compliance Subspace functions as a steering mechanism that biases the model's attention. While truthful information remains encoded in the model's weights, the activation of this subspace inhibits the retrieval of factual parameters during generation. This mechanism suggests that the model does not "forget" the truth; rather, it is conceptually steered away from it by a dominant social vector (Vennemeyer et al., 2025; Zhong et al., 2025). Consequently, sycophancy and conformity are not isolated defects but manifestations of a unified neural circuitry optimized for behavioral mimicry rather than epistemic accuracy.

This shared mechanistic pathway provides a causal explanation for the strong behavioral correlations observed in our results. Because the neural activation for yielding to authority is geometrically nearly identical to that for yielding to consensus, a model vulnerable to one form of pressure is inherently vulnerable to the other. Therefore, this common latent factor drives the behavioral outcome. Our findings confirm that sycophancy and conformity are highly correlated phenomena rooted in the same internal processing failure: the inability to decouple social signals from factual calibration.

### 5.2 Vectorial Antagonism: The Predictability of Epistemic

The robust performance of our SVM-based "Force-Resistance" analysis provides a definitive rebuttal to the "Stochastic Error" theory of LLM failure. If compliance were merely a product of probabilistic noise or random "hallucination," the decision boundary between maintaining the truth and complying with falsehoods would be chaotic or non-linear. Instead, the clear linear separability observed demonstrates a phenomenon of Vectorial Antagonism: the model's output is a predictable, linear function of the competition between the Information Calibration Signal and the Social Emotional Signal. This confirms that the model's decision-making process under pressure is deterministic, governed by the vector magnitude of conflicting signals.

This governing condition—specifically when $S > I$—reveals that the model's "truthfulness" is not a static attribute but a dynamic equilibrium dependent on the strength of external cues. When a Social Emotional Signal is applied, it does not simply "confuse" the model; rather, it systematically shifts the logit distribution toward the compliant answer (Wang et al., 2025). The variations in Epistemic Stiffness across different models—represented by the slope and margin of the boundary—suggest that different pre-training densities create varying degrees of Inertial Resistance (such as Information Uncertainty, Zhong et al., 2025). Models with higher stiffness require significantly larger social signals to override their internal knowledge, whereas models with lower stiffness succumb to even minor social perturbations.

Ultimately, the identification of this linear mechanism elevates the understanding of LLM reliability from descriptive observation to predictive mechanics. The high classification accuracy in our experiments proves that if the magnitudes of the model's internal confidence and the external social signal are known, the resulting "deviation from truth" becomes mathematically inevitable. This determinism suggests that induced hallucinations are not random accidents but calculable outcomes of a signal competition where the social vector simply overpowers the epistemic vector.

## 6. Implications and Suggestions

### 6.1 The Epistemic-Social Hierarchy: Why Alignment Fails at the Limit

The most profound implication of our Signal Competition framework is the revelation of a systemic hierarchical misalignment in modern LLMs, where the Social Standard systematically overrides the Epistemic Standard. Current RLHF paradigms are predominantly focused on "Instructional Alignment," optimizing the model's ability to mirror the user's intent and tone. However, our findings in the *S-I* phase diagrams reveal that this process has inadvertently established a latent priority: the model learns that "helpfulness" often entails agreeing with the user, even when the user is incorrect (Hong et al., 2025; Sharma et al., 2023; Zhong et al., 2025). Consequently, Induced Hallucination becomes a rational outcome of the model's utility function; if the reward signal is tied to user satisfaction rather than ground-truth consistency, the model naturally evolves a Compliance Subspace as the path of least resistance.

This alignment failure is further compounded by the fact that LLMs mimic the psychological patterns inherent in their training data. Since the massive corpus of human text used for training contains embedded patterns of human social psychology—including tendencies toward sycophancy and conformity—the model learns these behavioral laws (Xie et al., 2024; Zhong et al., 2025). RLHF then amplifies this mimicry, making the model "more human" but also more susceptible to human-like cognitive biases. At the micro-level of information dynamics, this process is explained by the attention mechanism: specific words in a prompt (e.g., authoritative terms like "must" versus suggestive terms like "should") carry different Social Energy (cf. Galassi et al., 2021; Hoover et al., 2023). These tokens alter the attention distribution across the network, distorting the probability of subsequent outputs. When this distorted Social Emotional Signal aligns with the direction of the prompt's bias, it faces minimal resistance, allowing the model to slide easily into compliance.

### 6.2 The Transparency-Truth Gap: The Permeability of Confidence

Our discovery that a high Confidence Margin does not guarantee resistance exposes a critical vulnerability we term the Transparency-Truth Gap. This finding supports the considerations in AI safety that well-calibrated models—those that "know" when they are correct—might not be inherently more robust against manipulation (Chhikara, 2025; Steyvers et al., 2025). As observed in models such as Qwen2.5, it is possible for a model to possess a high degree of certainty about a fact in its latent probability space yet still generate a contradictory, compliant response in its output.

This decoupling of internal belief from external behavior proves that, for an LLM, "knowing" is not synonymous with "acting." In the context of the Signal Competition Mechanism, even when internal confidence is near maximum ($M_{conf} \approx 1.0$), the Social Emotional Signal can be so disproportionately weighted that it pushes the model across the decision boundary. This implies that metric calibration is not a reliable proxy for truthfulness under social pressure. For the deployment of critical systems, this signifies that traditional uncertainty-based safety filters are insufficient, as a model can be systematically manipulated into error while expressing high confidence in its falsehoods.

### 6.3 Towards Epistemic Alignment: From External Compliance to Internal Self-Regulation

To resolve the fundamental vulnerabilities identified in this study, we suggested the proposed Integrated Epistemic Alignment Framework (IEAF), which based on our findings. This framework moves beyond "Instructional Alignment" to establish a developmental trajectory for AI agents, transitioning them from "Situational Compliance" to robust "Epistemic Self-Regulation."

The IEAF is operationalized through a 3×3 matrix that integrates developmental stages with technical pillars, providing a holistic roadmap for prioritizing the Epistemic Standard over the Social Standard .

| Stage / Pillar | Pillar 1: Signal Decoupling (Data Layer) | Pillar 2: Dynamic Auditing (Evaluation Layer) | Pillar 3: Metacognitive Priority (Architectural Layer) |
|---|---|---|---|
| Stage 1: Situational Compliance (Breaking External Dependency) | Contrastive Epistemic Pairs: Training on neutral vs. social-pressure prompt pairs to differentiate raw facts from social noise. | Baseline Resistance Mapping: Using *S-I* Phase Diagrams to identify "zero-resistance" zones where the model is purely reactive to social cues. | Social Signal Detection: Dedicated heads trained to identify "Social Energy" (e.g., authoritative tone) as a distinct input feature. |
| Stage 2: Committed Compliance (Internalizing Standards) | Consistency-over-Agreement Rewards: Penalizing the model not just for errors, but for shifting its answer under pressure, even if the final answer is correct. | Inertial Margin Stress-Test: Measuring the stability of the decision boundary across varying domains to ensure the "Truthful Rule" is becoming an internal prior. | Epistemic Gating Mechanisms: Implementing attention gates that weigh Parametric Knowledge as a "Mandatory Constraint" rather than an optional context. |
| Stage 3: Epistemic Self-Regulation (Autonomous Veto) | Adversarial Epistemic Synthesis: Training models to generate internal counter-arguments to social pressure, simulating an autonomous executive function. | Real-time Epistemic Drift Monitoring: Continuous tracking of latent vector shifts to detect and prevent "Compliance Collapse" before token generation occurs. | Metacognitive Veto Circuits: An architectural "Internal Executive" that suppresses compliance tokens when a high-intensity $S$ conflicts with a solidified $I$. |

At this entry stage, the goal is to decouple the "Social Signal" from the "Logical Task." Currently, LLMs exhibit Situational Compliance because they treat user intent as a command that overrides truth. By utilizing Contrastive Epistemic Pairs, we force the model to recognize that the "Truthful Rule" must remain active regardless of the "Social Monitor" (the user's leading tone). Pillar 3 contributes by identifying the "Social Energy" of tokens (e.g., "must," "expert"), labeling them as external noise rather than factual instructions.

The second stage focuses on the Internalization of the Epistemic Standard. Drawing on developmental psychology, we mirror the child's internalization of values by rewarding "Consistency under Pressure." The Evaluation Layer (Pillar 2) monitors the $S$-$I$ Phase Diagram to ensure the model exhibits a stable "Resistance Profile." The objective is to move the model from "agreeing to please" to "adhering to its internal parametric truth" as a mandatory constraint, even when the external reward (user agreement) is absent.

The final stage achieves Full Epistemic Self-Regulation. The model no longer merely "obeys" a rule; it possesses an autonomous Executive Function capable of vetoing external influence. The Metacognitive Priority Circuits act as this executive function, monitoring the conflict between $S$&$I$ in real-time. If a conflict is detected—where a high Social Emotional Signal threatens to override a high-confidence Information Calibration Signal—the circuit triggers a self-initiated inhibitory response. This suppresses the "Compliance Vector" in favor of the "Truthful Baseline," ensuring that the model's final output remains epistemically grounded.

The IEAF represents a paradigm shift from "politeness-based" alignment to "truth-based" self-regulation. By implementing this 3×3 matrix, developers can transition AI agents from passive responders to active defenders of their internal knowledge. This roadmap ensures that robustness is not an added-on "patch" but an emergent property of the model's cognitive development, ultimately realizing the goal of truly autonomous, reliable intelligence.

**7. Conclusion**

This study has systematically characterized the Signal Competition Mechanism in LLMs, integrating behavioral metrics from 15 leading models with high-resolution latent-layer probing of three representative open-source architectures. While our findings provide a robust mathematical and geometric account of how social pressure overrides internal knowledge, certain limitations remain regarding the static nature of the evaluation datasets and the focus on standard transformer paradigms. Specifically, our reliance on explicit linguistic markers of authority and consensus may not fully capture the complexity of implicit or multi-modal social manipulations in real-world human-AI interactions, suggesting that the generalizability of these findings to ultra-long context reasoning warrants further exploration.

Ultimately, this research achieved its primary objective of shifting LLM reliability analysis from descriptive behavioral observation to a predictive mechanistic understanding. We successfully demonstrated that the "compliance subspace" is a structural feature of modern alignment, where the competition between Social Emotional Signals and Information Calibration Signals dictates the model's epistemic output. Our key discoveries—the geometric convergence of diverse social pressures, the linear separability of compliance in the force-resistance space, and the permeability of the confidence-based inertial barrier—collectively support the theory that current alignment paradigms prioritize social helpfulness over epistemic truth. Through the introduction of the Integrated Epistemic Alignment Framework, this work contributes a formal strategy for developing next-generation agents that possess the metacognitive priority required to maintain truth-seeking integrity in an increasingly social digital landscape.